\documentclass{svproc}
\usepackage{url}
\usepackage{graphicx}

\begin{document}
\mainmatter              
\title{Measurement of azimuthal correlations between D mesons with charged particles in pp collisions at $\sqrt{s}$ = 7 TeV with ALICE}
\titlerunning{Heavy flavour correlations from ALICE}  
%
\author{Sonia Rajput, for the ALICE Collaboration}
\authorrunning{S.Rajput} 
%

%
\institute{University of Jammu, India\\
\email{sonia.rajput@cern.ch}\\ 
}

\maketitle              

\begin{abstract}
The azimuthal correlation of D mesons with charged particles in pp collisions at $\sqrt{s}$ = 7 TeV was measured  using the ALICE detector at LHC. This
measurement allows us to gain insight into heavy-flavour production processes and parton shower. The results are compared with the measurement performed in p-Pb collisions. The yield of charged particles in the near-side correlation peak and the peak width measured in the two collision systems are compatible, within the uncertainties, and described by Monte Carlo simulations.

\end{abstract}

\section{Introduction}
The goal of the ALICE experiment at the LHC is to study nuclear matter at extreme conditions of high temperature and high density at which quarks are no more confined into nucleons, giving rise to a new state of matter known as Quark Gluon Plasma (QGP)~\cite{ref:qgp}. Due to their large masses, heavy quarks (charm and beauty), are produced primarily in the early stages of heavy-ion collision, in hard partonic scattering processes, and experience the full evolution of the system propagating through the medium produced in such collisions. Therefore, they  are an effective probe to study the medium and provide information on its evolution. The analysis of angular correlations between heavy-flavour particles and charged particles serves as a tool  to characterize the heavy quark fragmentation process and its possible modification in the presence of the QGP. To this purpose, the measurement in pp collisions is necessary to obtain  a reference for the studies in heavy-ion collisions. The measurements in p-Pb collisions and their comparison with the pp measurements can give insight on how cold nuclear matter effects affect heavy-quark production and hadronization in p-Pb collisions.
\par
The ALICE  apparatus has excellent capabilities for heavy flavour measurements. A detailed description of the ALICE detector and its performance can be found in~\cite{ref:Alice}. D mesons and their charge conjugates are reconstructed in hadronic decay channels and selected exploiting the typical displaced-like topology of the decay vertices, PID (particle identification) and reconstruction quality cuts on the daughter tracks~\cite{ref:AliceCuts}. 

\section{D meson-charged particle angular correlation}

Two-dimensional angular correlations of D mesons ($\rm D^{0},D^{+},D^{*+}$) with charged particles are evaluated in pp collisions at $\sqrt{s}$ = 7 TeV for different ranges of D-meson and associated charged particle \textit{p}$\rm_{T}$. The contribution of background candidates is removed using the correlation distribution from the candidates in the sidebands of the D-meson invariant mass distribution. To account for the limited detector acceptance and detector spatial inhomogeneities, a correction factor is applied using the event mixing technique. Corrections for reconstruction and selection efficiency of D mesons and charged particles are also applied.  PYTHIA simulations~\cite{ref:Pythia} were used to produce template distributions of angular correlations between D mesons from B-meson decays and charged particles. These distributions, normalized to the expected amount of feed-down contribution evaluated from FONLL calculations~\cite{ref:Fonll} and considering the reconstruction efficiency of feed-down D mesons, are subtracted from the inclusive correlation distributions obtained from data. The fully corrected 2D correlations were projected onto the $\rm\Delta\phi$ axis, producing azimuthal correlation distributions normalized to the number of trigger D mesons. 
\begin{figure}
\centering
\includegraphics[width=9cm]{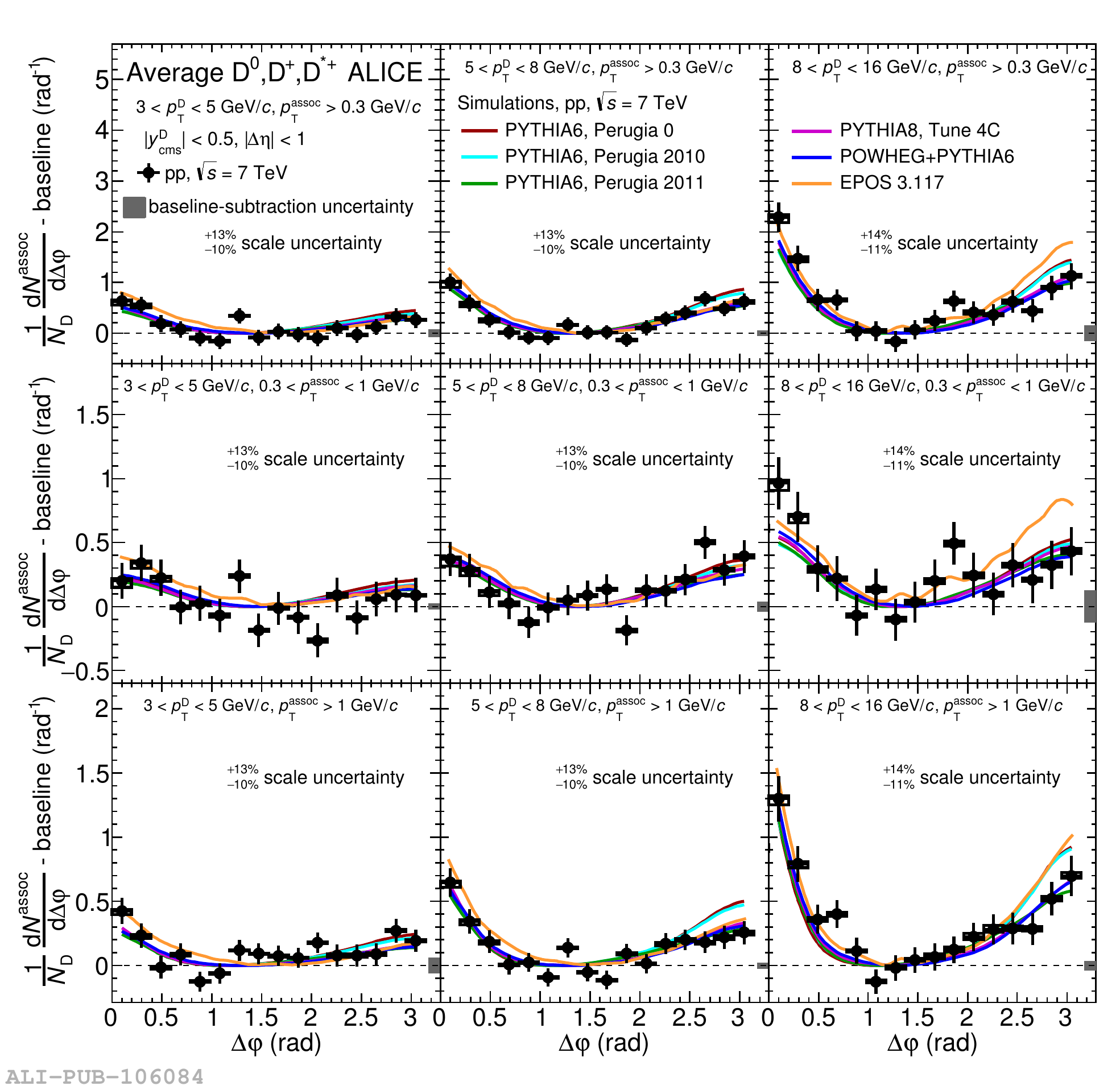}
\caption{Comparison of the baseline-subtracted D meson-charged particle azimuthal correlation distributions in pp collisions and predictions from different models for different ranges of D-meson and the associated charged particle \textit{p}$\rm_{T}$~\cite{ref:Paper}.}
\end{figure}

A weighted average of the results for the three D-meson species was then performed to reduce the statistical uncertainty. A fit with a function composed
of a constant term and two Gaussian functions modelling the near-side ($\rm\Delta\phi$ $\sim$ 0) and away-side correlation peaks was done to estimate the near-side peak associated yield and peak width and the height of the baseline.\\
The azimuthal correlation distributions in pp collisions are described well within uncertainties by expectations of various event generators (PYTHIA~\cite{ref:Pythia}, POWHEG~\cite{ref:Powheg} and EPOS~\cite{ref:epos}) in all kinematic ranges, as shown in Figure 1. Figure 2 shows the compatibility of the near-side yields measured in pp collisions at $\sqrt{s}$ = 7 TeV and p-Pb collisions at $\sqrt{s_{\rm{NN}}}$ = 5.02 TeV indicating that within the current uncertainties, no significant effect of cold nuclear matter effects in p-Pb collisions is observed from the data.

\begin{figure}
\centering
\includegraphics[width=8.5cm]{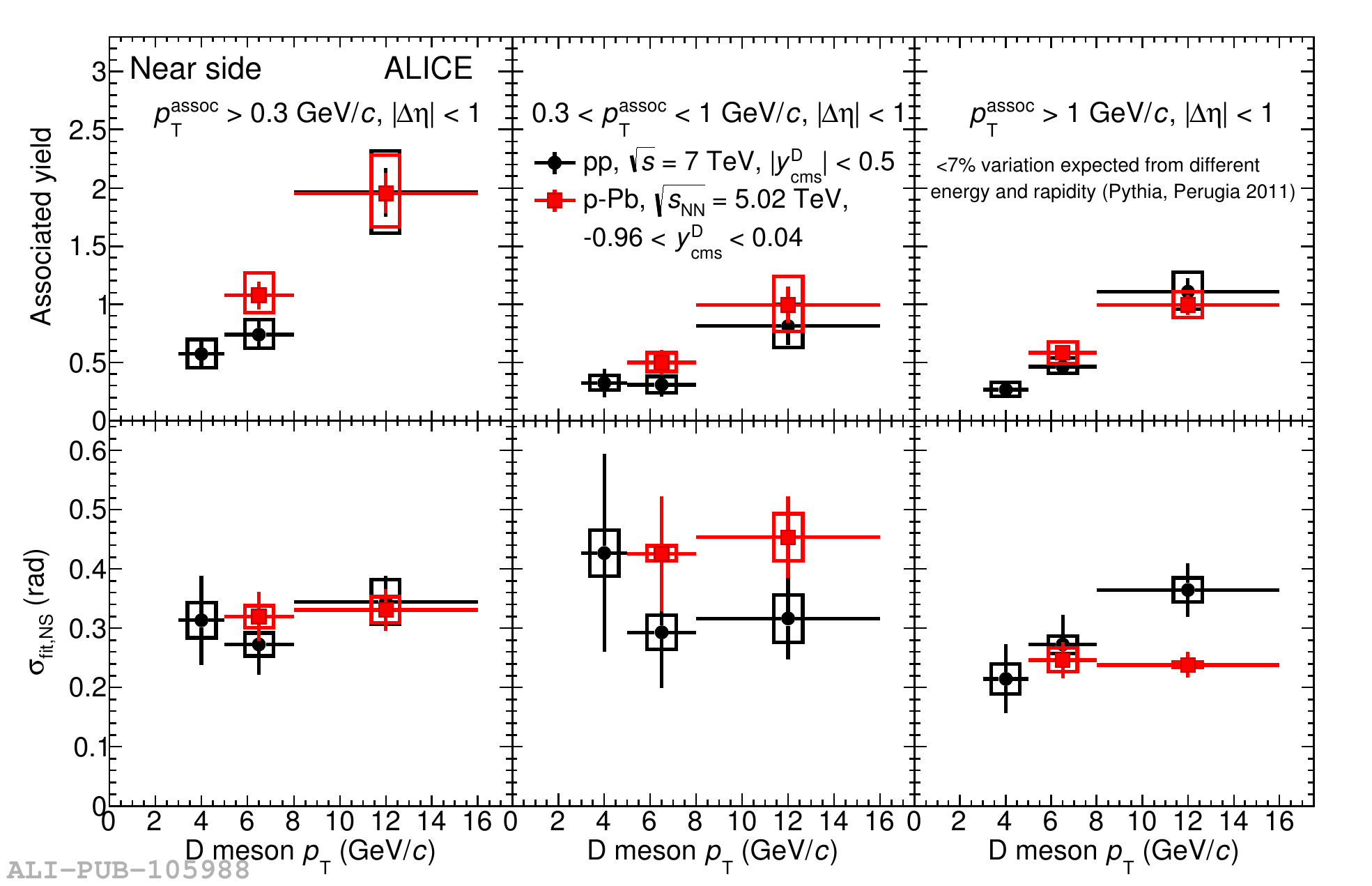}
\caption{Comparison of near-side associated yields (top) and near-side widths (bottom) extracted in pp and p-Pb collisions as a function of D-meson \textit{p}$\rm_{T}$~\cite{ref:Paper}. The near-side yields are extracted by integrating the near-side Gaussian component of the fit function.}
\end{figure}


%
%


\begin{thebibliography}{6}
%
\bibitem{ref:qgp} G.Brumfiel, What’s in a Name?, Nature, July 26(2004).
\bibitem{ref:Alice} ALICE Collaboration, Int. J. Mod. Phys. A29 (2014) 1430044.
\bibitem{ref:AliceCuts} ALICE Collaboration, JHEP 01 (2012) 128.
\bibitem{ref:Pythia} T.Sjostrand, S. Mrenna, P.Z. Skands, JHEP 05(2006) 26.
\bibitem{ref:Fonll} M. Cacciari, S. Frixione, N. Houdaeu, M.L. Mangano, P.Nason, G. Ridolfi, JHEP 1210 (2012) 137.
\bibitem{ref:Paper} ALICE Collaboration, ArXiv: 1605.06963.
\bibitem{ref:Powheg} S. Frixione, P. Nason, and C. Oleari, “Matching NLO QCD computations with Parton Shower simulations: the POWHEG method,” JHEP 11 (2007) 070.
\bibitem{ref:epos} K. Werner, B. Guiot, I. Karpenko, T. Pierog, "Analysing radial flow features in p-Pb and p-p collisions at several TeV by studying identified particle production in EPOS3", Phys. Rev. C89 (6)  (2014)  064903.
\end{thebibliography}
\end{document}